\documentclass{sf2a-conf2013}
\usepackage{graphicx}
\usepackage{hyperref}
\usepackage[]{natbib}  
\usepackage{epstopdf}

\def\BibTeX{{\rm B\kern-.05em{\sc i\kern-.025em b}\kern-.08em
    T\kern-.1667em\lower.7ex\hbox{E}\kern-.125emX}}
\bibpunct{(}{)}{;}{a}{}{,}  

\begin{document}

\TitreGlobal{SF2A 2013}


\title{Quasi-periodic oscillations from Rossby Wave Instability}

\runningtitle{QPO from Rossby instability}

\author{F. H. Vincent}\address{N. Copernicus Astronomical Centre, ul. Bartycka 18, 00-716, Warszawa, Poland}

\author{P. Varniere}\address{AstroParticule et Cosmologie (APC), Universit\'e Paris Diderot, 10 rue A. Domon et L. Duquet, 75205 Paris Cedex 13, France}

\author{H. M\'eheut}\address{CEA, Irfu, SAp, Centre de Saclay, F-91191 Gif-sur-Yvette, France}

\author{T. Paumard}\address{LESIA, Observatoire de Paris, CNRS, Universit\'e Pierre et Marie Curie, Universit\'e Paris Diderot, 5 place Jules Janssen, 92190
Meudon, France}

\author{G. T\"or\"ok}\address{Institute of Physics, Faculty of Philosophy and Science, Silesian University in Opava, Bezruÿcovo n‡m. 13, CZ-74601 Opava,
Czech Republic}

\author{M. Wildner$^5$}




\setcounter{page}{237}


\maketitle


\begin{abstract}
We study the Rossby wave instability model of high-frequency quasi-periodic
oscillations (QPO) of microquasars. We show ray-traced light curves of QPO within
this model and discuss perspectives of distinguishing alternative QPO models
with the future Large Observatory For X-ray Timing (LOFT) observations.
\end{abstract}



\section{Introduction}

High-frequency quasi-periodic oscillations (QPOs) of microquasars have attracted a lot of attention
in the recent years~\citep[for a review see][]{remillard06}. Up to now, no consensus has emerged
regarding the nature of these events.
Many different models have been proposed, that are still today candidates to account
for QPOs.
\citet{ste-vie:1999} and \citet{ste-etal:1999} suggest that QPOs could arise from the modulation of the X-ray flux by the periastron precession and the Keplerian frequency of blobs of matter orbiting in
an accretion disk around the central compact object.
Also QPOs could be due to modulation of the X-ray flux by oscillations 
of a thin accretion disk surrounding the 
central compact object~\citep[see][]{wagoner99,kato01}.
\citet{fragile01} propose that QPOs are due to the modulation
of the X-ray flux caused by a warped accretion disk surrounding
the central compact object.
Pointing out the 3:2 ratio of some QPOs in different sources,
\citet{abramowicz01} have proposed a resonance model
in which these pairs of QPOs are due to the beat between the
Keplerian and epicyclic frequencies of a particle orbiting
around the central compact object.
\citet{schnittman04} investigated in great detail predictions of a model of hot spot radiating isotropically on nearly circular equatorial orbits. In their study, the hypothetic resonance between the Keplerian and radial epicyclic frequencies gives rise to peaks in the modeled power spectrum.

\citet{tagger06} advocate the fact that QPOs in microquasars could
be triggered by a Rossby wave instability (RWI) in the accretion
disk surrounding the central compact object. Ray-traced light curves
have been recently developed for this model by~\citet{vincent13}. 
This article aims at exploring the case for the RWI model of QPOs,
as well as discussing perspectives of distinguishing alternative
models with future observations.
  
\section{Scenario of a quasi-periodic oscillation within the Rossby model}

\begin{figure}[ht!]
 \centering
 \includegraphics[width=0.48\textwidth,clip]{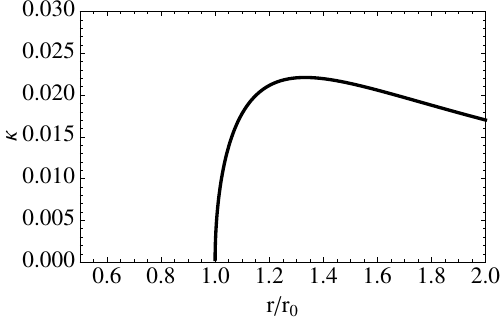} \hspace{0.5cm}   
 \includegraphics[width=0.4\textwidth,clip]{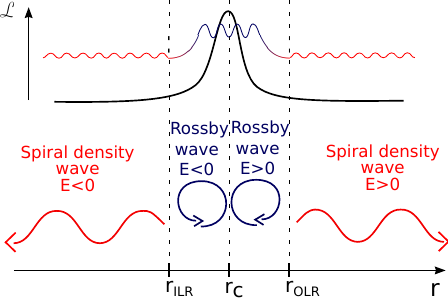} 
  \includegraphics[width=0.4\textwidth,clip]{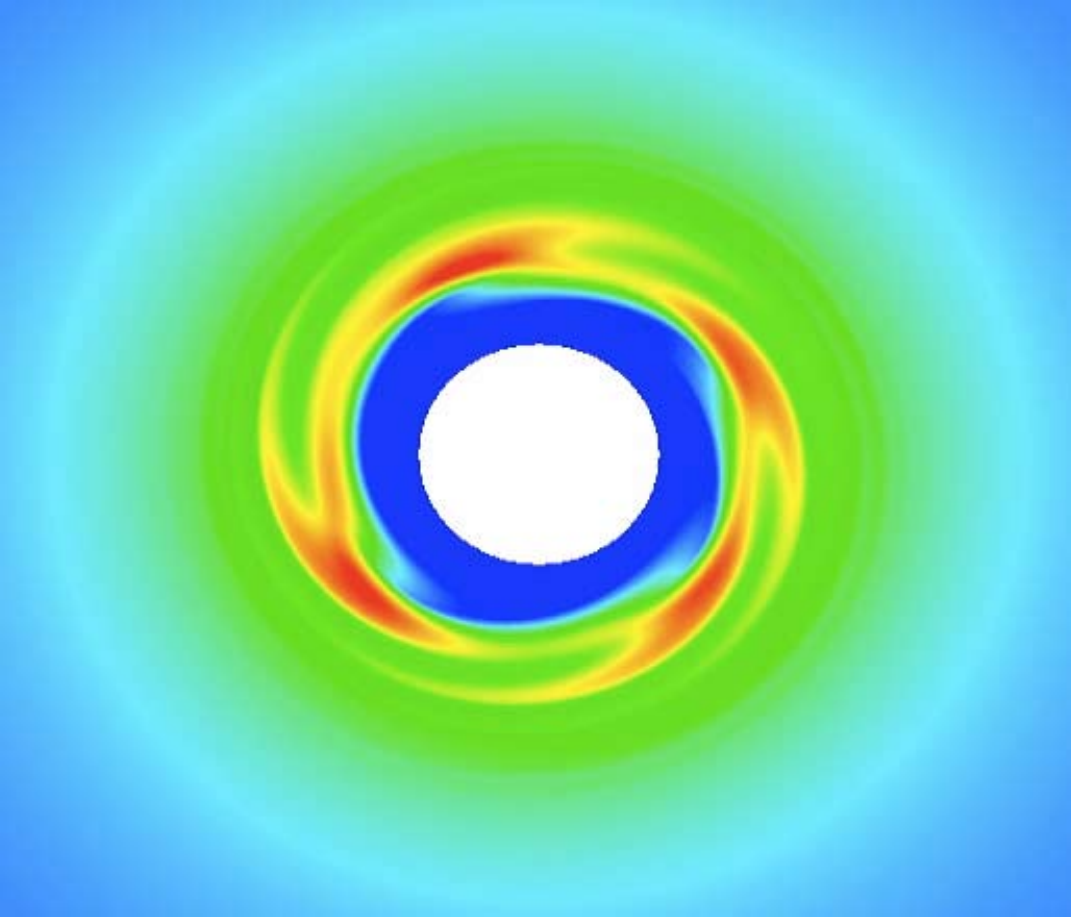}      
  \caption{{\bf Upper left:} radial epicyclic frequency distribution $\kappa$ as a function of $r$ in an accretion disk
  	surrounding a Schwarzschild black hole. An extremum occurs at a radius somewhat bigger than the ISCO radius $r_0$. 
	{\bf Upper right:} development of the Rossby wave instability. Density waves propagate on both sides of the extremum
	(or corotation) radius $r_C$, and Rossby vortices appear at the corotation~\citep[figure from][]{meheut13}. However, in our
	case where the instability is triggered near ISCO by the extremum of epicyclic frequency, hardly any waves are emitted at radii
	smaller than the corotation radius. {\bf Lower:} density map of the disk when the
	Rossby instability is fully developed. Four spiral arms are clearly visible while most of the density is concentrated
	in the four Rossby vortices.}
  \label{fig:launch}
\end{figure}

The Rossby wave instability~\citep{lovelace99} will be triggered in an accretion disk
provided the following quantity

\begin{equation}
\mathcal{L} = \frac{\Sigma \Omega}{2 \kappa^2} \frac{p}{\Sigma^\gamma},
\end{equation}

where $\Sigma$ is the surface density, $\Omega$ the rotation velocity,
$\kappa$ the radial epicyclic frequency, $p$ the pressure and $\gamma$
the adiabatic index, reaches an extremum. For an accretion disk surrounding a Schwarzschild black hole,
this quantity will naturally reach an extremum as the epicyclic frequency reaches
a maximum close to the innermost stable circular orbit (ISCO) as illustrated
in Fig.~\ref{fig:launch}, in the upper left panel.  

\begin{figure}[ht!]
 \centering
 \includegraphics[width=0.48\textwidth,clip]{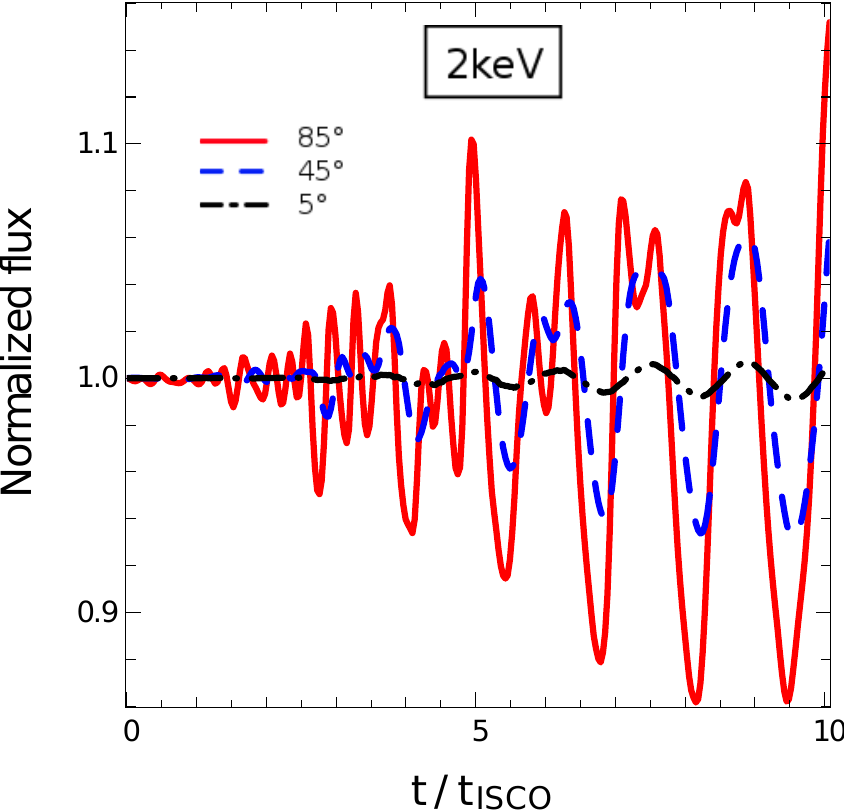}%
  \caption{Ray-traced light curve of an accretion disk surrounding a Schwarzschild black hole subject to the Rossby wave instability,
  	seen under an inclination of $5^{\circ}$ (black - face-on), $45^{\circ}$ (blue) or $85^{\circ}$ (red - edge-on). The energy of the
	observed radiation is $2$~keV. The time is given in units of the ISCO Keplerian orbital time.
  	The modulation of the intensity due to relativistic beaming leads to a modulation of the light curve at a few
	percent level, highly depending on the inclination. At high inclination, the modal signature of the instability appears clearly
	with higher mode number at early times, decreasing to mode $m=1$ dominating at the end of the simulation.}
  \label{fig:LC}
\end{figure}

Once the instability is triggered, it will develop spiral density waves that will spread
on both sides of the extremum (or corotation) radius (see Fig.~\ref{fig:launch}, upper right 
and lower panel). Rossby vortices develop at the corotation radius and aggregate most
of the density. The mode number of the instability, or number of spiral arms ($m=4$
for instance in the lower panel of Fig.~\ref{fig:launch}) is evolving with time, always
from higher values of $m$ to smaller values of $m$. When the instability is fully developed,
modes $m=1$ or $2$ or $3$ can dominate, or a combination thereof.
At a given time, different modes
can coexist. This is clearly visible in Fig.~\ref{fig:LC} showing the ray-traced light curve
of an accretion disk surrounding a Schwarzschild black hole subject to the RWI: the red
curve clearly shows the evolution, and superimposition, of different modes.

To summarize, here is one possible scenario that would lead to QPOs due
to the RWI in microquasars:

\begin{itemize}
\item initial state: an accretion disk around a Schwarzschild black hole with inner radius $r_{\mathrm{in}}$
	above the radius of the maximum of epicyclic frequency $\kappa$; no instability;
\item when $r_{\mathrm{in}}$ becomes smaller than the maximum of $\kappa$, the RWI is triggered
	and develops, with different modes being present; this is the QPO;
\item as $r_{\mathrm{in}}$ increases, it gets bigger than the maximum of $\kappa$
	and the instability is quenched. 
\end{itemize}

As a consequence, the evolution of the inner radius of the accretion disk dictates the
triggering of the instability and the appearance of QPOs. 

\section{Towards distinguishing alternative models?}

One very interesting feature of the RWI model is the modal signature that is clearly 
illustrated in Fig.~\ref{fig:LC}. 

Our future aim is to develop observational strategies to distinguish alternative
models of QPOs. In this perspective the future X-ray timing instrument
LOFT~\citep{feroci12} is very interesting, as its spectral ability allows
it to generate much more precise spectra than the current RXTE data.
Future work will thus be devoted to simulating LOFT power spectra of
QPOs described by various models, in order to determine whether the
spectral signature of the various models may be used to distinguish them.

\begin{figure}[ht!]
 \centering
 \includegraphics[width=\textwidth,clip]{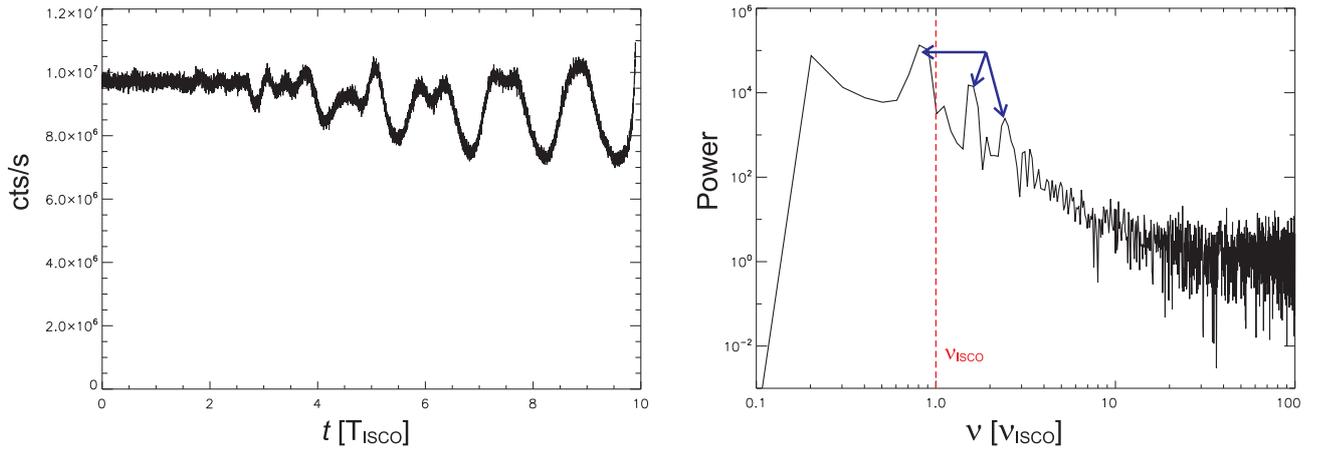}%
  \caption{{\bf Left:} simulated light curve of a LOFT observation of RWI in an accretion disk surrounding a $1000\,M_{\odot}$ black hole (see text for details).
  	 {\bf Right:} corresponding power spectrum. The spectral signature of the RWI, with three peaks at one, two and three times a fundamental frequency,
	 is clearly visible. }
  \label{fig:LOFT}
\end{figure}

Fig.~\ref{fig:LOFT} gives a very first, illustrative, LOFT simulation. It shows a light curve
and power spectrum, as observed by LOFT, corresponding to the theoretical light curve depicted in Fig.~\ref{fig:LC}.
However, this simulation assumes a black hole of the order of $1000 \,M_{\odot}$, which changes
the time scale and allowed us to get this illustrative result without having to resort to intensive
computations which will be needed to reach the time resolution required for a $10 \,M_{\odot}$ black hole,
typical of a microquasar. Fig.~\ref{fig:LOFT} must then be seen as an illustration of the
simulations we are aiming at. It shows very clearly the spectral signature of the RWI with three
distinct peaks at one, two and three times a fundamental frequency a bit lower than the ISCO
frequency. This is logical as the RWI develops at a radius somewhat bigger than the ISCO radius.

\section{Conclusions}
We have advocated here one model of high-frequency QPOs of microquasars: the Rossby wave instability model.

This model allows to modulate the light curve of microquasars at a few percent level, with the modulation
depending strongly on inclination. It allows to explain naturally the coexistence of different modes in the
observed QPO signal.

Future work will be dedicated to developing LOFT simulations of this RWI model, and compare its
observed spectral signature to alternative models.

\begin{acknowledgements}
This work has been financially supported by the French
Programme National des Hautes Energies (PNHE).
Some of the simulations were
performed using HPC resources from GENCI-CINES (Grant 2012046810).
\end{acknowledgements}

\bibliographystyle{aa}  
\bibliography{vincent} 

\end{document}